\documentstyle[11pt,epsf]{article}
\def\beq{\begin{equation}}
\def\enq{\end{equation}}
\def\eeq{\end{equation}}

\def\ww{{\cal W}}
\def\ww3{{\cal W}^{3}}

\def\tev{{\rm \, TeV}}
\def\gev{{\rm \, GeV}}

\def\ra{\rightarrow}
\def\bar{\overline}

\def\slashchar#1{\setbox0=\hbox{$#1$}           
   \dimen0=\wd0                                 
   \setbox1=\hbox{/} \dimen1=\wd1               
   \ifdim\dimen0>\dimen1                        
      \rlap{\hbox to \dimen0{\hfil/\hfil}}      
      #1                                        
   \else                                        
      \rlap{\hbox to \dimen1{\hfil$#1$\hfil}}   
      /                                         
   \fi}                                         %
\def\centeron#1#2{{\setbox0=\hbox{#1}\setbox1=\hbox{#2}\ifdim
\wd1>\wd0\kern.5\wd1\kern-.5\wd0\fi
\copy0\kern-.5\wd0\kern-.5\wd1\copy1\ifdim\wd0>\wd1
\kern.5\wd0\kern-.5\wd1\fi}}
\def\ltap{\;\centeron{\raise.35ex\hbox{$<$}}{\lower.65ex\hbox{$\sim$}}\;}
\def\gtap{\;\centeron{\raise.35ex\hbox{$>$}}{\lower.65ex\hbox{$\sim$}}\;}
\def\gsim{\mathrel{\gtap}}
\def\lsim{\mathrel{\ltap}}
\def\D0{D\O}
\def\doublespaced{\baselineskip=\normalbaselineskip\multiply
    \baselineskip by 150\divide\baselineskip by 100}
\def\singlespaced{\baselineskip=\normalbaselineskip}
\def\NPB#1#2#3{Nucl. Phys. {\bf B#1} (19#2) #3}
\def\PLB#1#2#3{Phys. Lett. {\bf B#1} (19#2) #3}
\def\PLBold#1#2#3{Phys. Lett. {\bf #1B} (19#2) #3}
\def\PRD#1#2#3{Phys. Rev. {\bf D#1} (19#2) #3}
\def\PRL#1#2#3{Phys. Rev. Lett. {\bf #1} (19#2) #3}
\def\PREP#1#2#3{Phys. Rep. {\bf #1} (19#2) #3}
\def\ZPC#1#2#3{Z.~Phys. {\bf C#1} (19#2) #3}
\def\charI{ { \tilde \chi}^{\pm}_1 }
\def\charIinv{ { \tilde \chi}^{\mp}_1 }
\def\charIplus{ {\tilde \chi}^{+}_1 }
\def\neutI{ { \tilde \chi}^0_1 }
\def\neutII{ { \tilde \chi}^0_2 }
\def\ltilde{ {\tilde \ell} }

\def\lL{ { \tilde \ell}_L }
\def\lLstar{ { \tilde \ell}_L^* }
\def\lR{ { \tilde \ell}_R }
\def\lRstar{ { \tilde \ell}_R^* }
\def\snu{ { \tilde \nu}_L }

\def\nubar{ \overline{ \nu } }

\def\h{ { h^0 } }
\def\Et{ { \slashchar{E}_T } }
\def\Etcut{ { \slashchar{E}_T^{\rm cut} } }
\def\sigbreff{ \sigma \times {\rm BR} \times {\rm EFF} }
\doublespaced
\begin{document}
\begin{titlepage}
\begin{flushright}
{\large
 hep-ph/9505245 \\
 CIT 68-1986 \\
 UM-TH-95-14 \\
 May 1995 \\
}
\end{flushright}
\vskip 2cm
\begin{center}
{\Large\bf
Possible Signals of Constrained Minimal Supersymmetry } \\
\vskip 4pt
{\Large\bf
at a High Luminosity Fermilab Tevatron Collider}
\vskip 1cm
{\large
Stephen Mrenna\footnote{{\tt mrenna@cithex.cithep.caltech.edu}} } \\
\vskip 8pt
{\it California Institute of Technology, Pasadena, CA 91125, USA }
\vskip 1cm
{\large
 G.~L.~Kane\footnote{{\tt gkane@umich.edu}},
 Graham D.~Kribs\footnote{{\tt kribs@umich.edu}},
 James D.~Wells\footnote{{\tt jwells@walden.physics.lsa.umich.edu}} } \\
\vskip 8pt
{\it Randall Physics Laboratory, University of Michigan,\\
     Ann Arbor, MI 48109--1120, USA } \\

\end{center}

\vskip .5cm

\begin{abstract}

We study the most promising signals of Constrained Minimal
Supersymmetry detectable at a luminosity upgraded 2 TeV Fermilab
Tevatron collider.  Using a full event--level Monte Carlo based on
{\sc Pythia/Jetset}, we simulate the trilepton signal
examining in detail the effect of constraints on the parameter space.
We also simulate the monolepton and dilepton signals,
the $\Et$ + jets signal, and the signals of stop production
in supersymmetry all with full Standard Model backgrounds
with realistic detector cuts.  We find that large fractions of
parameter space can be probed (or eliminated if no signal is found),
but mass limits on charginos and
neutralinos are not possible based solely on the trilepton signal.
Detection efficiencies depend strongly on supersymmetry parameters
beyond simply the neutralino and chargino masses; analyses (experimental
or theoretical) that do not include this will draw misleading
conclusions.  Finally, we comment on how searches at LEP~II
will complement searches at Fermilab.
\end{abstract}

\end{titlepage}
\setcounter{footnote}{0}
\setcounter{page}{2}
\setcounter{section}{0}
\setcounter{subsection}{0}
\setcounter{subsubsection}{0}

\doublespaced

\section{Introduction}
\indent

The search for supersymmetry (SUSY) should be one of the primary goals at the
Fermilab Tevatron and
any future collider planned or in construction.  While $e^+e^-$ colliders
generally have the advantage of unambiguously finding or excluding
superpartners up to $m \sim \frac{\sqrt{s}}{2}$, their mass reach
is strictly energy limited.  LEP~II, the highest energy
$e^+e^-$ collider approved, will be able to probe chargino masses up
to about $m_W$ but no further.  A useful complement to LEP~II's
important contributions to the search for supersymmetry is provided
by a luminosity upgraded Fermilab Tevatron $p\bar p$ collider
at $\sqrt{s}=2\tev$\@.  As we shall subsequently describe, if the correct
supersymmetric theory has a chargino mass below $m_W$, the upgraded
Tevatron might possibly fail to discover it (hence the importance
of LEP~II); however, a high luminosity Tevatron does have the
capability of discovering supersymmetry at mass scales {\em far exceeding\/}
the capabilities of LEP~II.

In this paper we report the results of our simulations of many
supersymmetric signals for a $p\bar p$ collider at $\sqrt{s}=2\tev$.
To this end, we introduce an implementation of all tree level
MSSM processes and decay modes in the event generators {\sc Pythia} and
{\sc Jetset}~\cite{sjostrand94:74,PYTHIA,JETSET}.
This implementation of the MSSM
is not only a cross check of ISASUSY~\cite{ISASUSY}
(see also~\cite{SUSYSM}), which uses the ISAJET~\cite{ISAJET}
generator, but includes the refinements inherent to the
{\sc Pythia/Jetset} system.  The full details of the event
generator will be published in a forthcoming technical
report~\cite{Mrenna}.

An equally important task is to run the simulations on
supersymmetric solutions which could be the correct theory of nature.
Therefore, we have simulated supersymmetric events in the context
of the Constrained Minimal
Supersymmetric Standard Model (CMSSM)~\cite{kane94:6173}.  Briefly, the CMSSM
enforces gauge coupling unification at the ``unification''
scale $\sim$$10^{16}\gev$,
enforces proper electroweak symmetry breaking, assumes common
scalar masses ($m_0$), common gaugino masses ($m_{1/2}$), and common trilinear
scalar soft couplings ($A_0$) at the unification scale,
requires R--parity
conservation, and imposes all known
experimental constraints such as limits on
$b\rightarrow s\gamma$ decays, invisible width constraints on
$Z\to \tilde\chi^0_i \tilde \chi^0_j$ at LEP, etc.
Each supersymmetric ``solution'', with
all of its masses and mixings, is determined by five input parameters
\begin{equation}
m_0, m_{1/2}, A_0, \tan \beta,~{\rm and}~{\rm sign}(\mu ).
\end{equation}
(We use $m_t = 170$ GeV\@.)
Our convention is that $\tan\beta =v_u/v_d$, where $v_u$ ($v_d$) is
the vev which gives mass to the up--type (down--type) fermions.
Our convention for the relative sign of $\mu$
is reflected by our choice of the
chargino mass matrix element $X_{22} = -\mu$ and the
neutralino mass matrix elements $Y_{34} = Y_{43} = +\mu$ (which
is the opposite convention from Haber and Kane~\cite{HaberKane}).

The requirements we impose on our supersymmetric solutions
are consistent and interrelated.  While it is not
absolutely necessary for nature to follow all of our
theoretical assumptions (such as common scalar masses),
we do wish to emphasize that each requirement supports
the other requirements to some degree.
For example, the experimental requirements such
as limits on flavor changing neutral currents support,
although do not absolutely require, the theoretical
preference in supergravity that all scalars have a common mass at
the high scale.  Similarly,
gauge coupling unification supports the notion of R--parity
conservation~\cite{martin92:2769,diehl95:399},
implying an absolutely stable lightest
supersymmetric particle (LSP), which is the lightest
neutralino $\neutI$.  Proper electroweak symmetry
breaking, which is meticulously enforced in the CMSSM, gives the LSP the
right properties to be a natural weakly interacting
cold dark matter particle, explaining many astrophysical
{\em observations\/}~\cite{sikivie95:292}.

One of the strengths of the CMSSM is that it is strongly constrained
and theoretically restrictive.  We believe it is remarkable
progress that it is possible to construct consistent supersymmetric models
that incorporate all the above constraints and provide rich
phenomenological predictions.  In general it is always appropriate
to study the simplest (hence minimal) theory that is consistent
with what is known from the Standard Model (SM), while simultaneously
extending our understanding of nature.  The CMSSM is minimal in the
sense that it has the spectrum and group structure of the SM plus
superpartners while implementing the minimal supergravity boundary
conditions at the unification scale.

Using the CMSSM framework we have performed {\em event\/}--level
analyses of the chargino--neutralino trilepton
signal, the chargino--chargino and slepton--slepton dilepton signals,
the slepton--sneutrino and chargino--LSP monolepton signals,
the squark/gaugino $\Et$ + jets signal and the signals from stop
production and top decays to stop assuming
various integrated luminosity scenarios
($200~{\rm pb^{-1}}$, $2~{\rm fb^{-1}}$ and $25~{\rm fb^{-1}}$)
for an upgraded Fermilab Tevatron collider.
There have been several previous studies of supersymmetric
signals~\cite{frere83:331,baer95:9503479} all of which are
useful and demonstrate the possibility of detecting supersymmetric
signals at Fermilab, but often make unrealistic assumptions about
supersymmetry or make simplifications in the simulation of the
signal and background.  This study is more general; we find
several results different from other studies which stem
from two basic principles unique to our approach.  First, we
only consider values of SUSY parameters which are consistent with
the CMSSM\@.  Second, within this general framework, we examine all
parameters rather than a set of special cases.

It is possible to construct supersymmetric frameworks that
modify some of the theoretical assumptions and this may be
necessary in the future.  However, since we sample
the supersymmetric parameter space with thousands
of solutions, it is likely that many alternatives will lie in
regions we have already covered in this study.  Hence, the
conclusions of this paper are not likely to be significantly
modified.  In fact, there are already some hints from LEP data
that suggest going beyond the minimality of the CMSSM is necessary
due to the $R_b$ and $\alpha_s$
measurements~\cite{wells94:219,shifman95:605,kane95:um-th-95-16}.
If these measurements are manifestations of non-minimal supersymmetry,
then we expect very light charginos and stops to exist, and
therefore the Fermilab Tevatron will have a much greater chance
to discover them.

We organize this paper as follows.  In Section 2 we briefly explain
the event simulation needed for every allowed solution in the CMSSM
parameter space.  Section 3 contains the motivation and results for
each signal we have studied.  We perform a full background analysis
on every signal, specifying the cuts needed to reduce backgrounds.  From
the background estimates we determine the detectability or significance
of each signal for particular integrated luminosities.
In Section 3.1 we examine monolepton and dilepton signals,
in Section 3.2 we examine the trilepton signal and
in Section 3.3 we examine in detail the effect of constraints
on supersymmetric parameter space with trilepton detection as an
example.  In Section 3.4 we examine the $\Et$ + jets signal,
in Section 3.5 we examine the signals from stop production and
in Section 3.6 we comment on the signal from top decays to stop.
In Section 4 we discuss going beyond the CMSSM, including possible
effects on supersymmetric signals and detection.
Finally, in Section 5 we conclude with a summary
of our results and a brief discussion of how a high luminosity
Fermilab Tevatron would complement LEP~II.

\section{Monte Carlo Simulation}
\indent

Each solution defined by the supersymmetric parameters
$m_0$, $m_{1/2}$, $A_0$, $\tan \beta$, ${\rm sign}(\mu )$ is
considered a complete potential theory of nature with well-defined
low energy (weak scale) masses, couplings, etc., obtained
from running the masses and couplings from the unification
scale to the weak scale.  All of these low energy parameters
must be set for each solution.  Then, events are generated
using 2--to--2 cross section
formulae~\cite{dawson85:1581,bartl86:441,bartl86:1}
incorporated into {\sc Pythia},  which also generates \
initial and final state QCD and QED radiation.
Each 2--to--2 process is derived from a $p{\overline p}$
collision using the CTEQ2L structure functions.
The sparticle decays are based on 2--body and 3--body formulae
for the decay rates~\cite{bartl86:441,bartl86:1,gunion88:2515}
added to {\sc Jetset}, which performs
string fragmentation and hadronization.  The final output of the event
generator is a list of ``stable'' particles and their 4-momenta.
These particles are then fed into a model detector which smears momenta
based on Gaussian energy resolution functions and defines jets based on
$E_T$ towers with $\phi\times\eta$ segmentation.
The detector--level variables are
used to define the experimental quantities upon which kinematic cuts are based.

Our model detector is based loosely on the CDF detector~\cite{CDF95:2626}.
The calorimeter is segmented
$\Delta\phi\times\Delta\eta = 0.1 \times 0.1$ with
$\eta$ coverage to $\eta = 4.2$.
The hadronic energy resolution is chosen so that the jet
resolution is roughly $0.7/\sqrt{E}$ with a degradation to $1.4/\sqrt{E}$
for large $\eta$.  The electromagnetic energy resolution is $0.2/\sqrt{E}$,
and the muon momentum resolution is $\frac{\sigma_{p_T}}{p_T} =
\sqrt{(0.0009p_T)^2+(0.0066)^2}$.  Electrons, muons, and jets are identified
for $\eta<$ 2.5, though the whole calorimeter is used to define the
$\slashchar{\vec{E}}_T$ vector.
Isolated leptons $(l)$ are
defined as electrons and muons with $E_T^{extra} = \sum_{i}^{} E_T^{(i)} -
E_T^{(l)} <$ 2 GeV within a cone
$R \equiv \sqrt{\Delta \phi^2+\Delta \eta^2} \leq 0.4$ around the lepton.
The sum $i$ is over all leptons, photons, and hadrons.
Jets $j$ are defined with $R = 0.6$ and $E_T^j > 15$ GeV.

Every signal must go through a process where thousands
of supersymmetric solutions are simulated each with thousands of events.
The final result is a set of observables (cross sections, decay rates)
which we analyze in the form of scatter plots with each point
representing a possible supersymmetric theory of nature.

\section{Supersymmetric Signals and Backgrounds}
\indent

We consider several potential signals and backgrounds for the
Fermilab Tevatron $p\overline{p}$ collider at $\sqrt{s}$~=~2~TeV\@
in the following sections organized by the particular
detection signature.  In Section 3.3 we have performed a
detailed analysis of the effect of constraints on the
supersymmetric parameter space with the trilepton signal
as our example.

\subsection{Monolepton and Dilepton Signals and Backgrounds}
\indent

We have simulated the monolepton signals from $\lL \snu$,
$\charI \neutI$ production
and the dilepton signals from $\charI \charIinv$, $\neutII \neutI$,
$\lL \lLstar$ and $\lR \lRstar$
production~\cite{baer89:303,baer94:3283,baer95:2159,lopez94:9412346,
baer95:9504234}.
Of course any monolepton
signal must compete against the huge background from
$W^\pm(\ra\ell^\pm\nu)$ production which is ${\sim}1$ nb.
For the supersymmetric
signals, the energy spectrum of the lepton from the
2--body decay of the $\ltilde$ or
the (2--body or 3--body) decay of the $\charI$ is generally soft because most
of the energy is carried by the superpartner decay product $\neutI$.
In addition, the classic $\Et$ signature for SUSY does not easily
distinguish signal from background.  In the end, we could
find no set of straightforward cuts which could reliably extract
a monolepton SUSY signal at the Fermilab Tevatron at any luminosity.

The dilepton signals appear much more promising, since
the large background from
$\gamma^*(\ra \ell^+\ell^-)$ and $Z(\ra \ell^+\ell^-)$
can be significantly
reduced with an $\Et$ cut and an invariant mass cut on opposite sign
leptons near the $Z$ mass.  However, the background from the smaller
$W^\pm(\ra\ell^\pm\nu) W^\mp(\ra\ell^\mp\nu)$ production is virtually
irreducible at a total leptonic cross section of 550~fb (without any cuts).
We applied a minimal set of cuts
and found the dilepton signals from $\neutII \neutI$, $\lL \lLstar$
and $\lR \lRstar$ are extremely difficult to pull out of
the $W^\pm W^\mp$ background of roughly $\sim$140 fb (with our cuts).
Only the $\charI \charIinv$ signal is visible in a small number
of solutions with a chargino mass reach of $m_{\charI}$ $\sim$80 GeV,
$\sim$110 GeV and $\sim$130 GeV for integrated luminosities
of 200 ${\rm pb}^{-1}$, 2 ${\rm fb}^{-1}$ and 25 ${\rm fb}^{-1}$.
No limits on chargino masses could be extracted even at an integrated
luminosity of 25 ${\rm fb}^{-1}$.  It is possible that some of the
solutions with detectable trilepton signals described
in the next section could be confirmed by the dilepton
signal~\cite{baer95:9504234}.

\subsection{Trilepton Signal and Backgrounds}
\indent

The trilepton signal from $\charI\neutII$ production is probably the
most promising signal of supersymmetry at a hadron
collider~\cite{trileptons,kamon94:19,baer95:2159,
lopez94:9412346,baer95:9504234}.
Although the signal is obtained from the chargino decaying to one lepton
and the neutralino decaying to two leptons as in the previous section,
the resulting trilepton signature has two clear advantages over the
monolepton and dilepton signals.  First, the correlations induced
between the mass
and mixing parameters in the CMSSM almost invariably output a lightest
chargino which is mostly
a charged wino ($\tilde W^\pm$) and a second lightest neutralino which
is largely a neutral wino ($\tilde W^3$).  Thus, the coupling at the
$W^\mp{\tilde \chi}^\pm_1{\tilde \chi}^0$ vertex is near maximal
in the CMSSM since it
is the supersymmetrized version of the $W^\pm W^\mp W^3$ Standard Model
vertex.  Second and most important,
the trilepton signal has few sizeable backgrounds in stark contrast to
the monolepton and dilepton signals.  In this section we explore
all aspects of the trilepton signal including a full background analysis
and a complete examination of the CMSSM parameter space.

There are six sources of physics\footnote{Detectors have additional
backgrounds from, for example, $\gamma$'s that ``fake'' an electron.
The dominant background where this can occur is for
$\gamma^*/Z \ra \tau^+\tau^- \ra$ e's/$\mu$'s with a $\gamma$ (and $\Et$)
in the final state.  We find that at a ``fake'' rejection rate
$\lsim 10^{-4}$ our set of cuts effectively eliminates this background.}
backgrounds relevant to the trilepton signal:  $WZ$,
$ZZ$, $t\overline{t}$, $t\overline{b}+b\overline{t}$,
$Z + g$ and $W + g$.  The $WZ$, $ZZ$ and $t \overline t$ backgrounds
have been studied before
(see {\it e.g.}~Refs.~\cite{kamon94:19,baer95:2159}).
Our results are similar, but more general since we
study both the traditional backgrounds (first three) and
also backgrounds from the leptonic decays of $b$ quarks including
those from $g^*\rightarrow b\overline{b}$ splitting.  The latter
can be important especially if one considers measuring leptons with low
$p_T \approx 5$ GeV, as we do in this study.  For the $WZ$, $ZZ$ and $Z+g$
backgrounds, we properly incorporate the effect of
$\gamma^*$ interference in $Z$ production
and use the full $2 \ra 4$ matrix elements inherent in the {\sc Pythia}
Monte Carlo generator.  We also find it is essential to simulate
all possible decay channels of gauge bosons that could possibly
lead to leptons in the final state ({\it i.e.},~we include the tau decay
channels).  In the end, we find the $WZ$ and $ZZ$
backgrounds are the most difficult to eliminate, since they contain
real isolated leptons (distinguishable from leptons from heavy quark
decay which are generally not isolated from hadrons).

Below, we enumerate the kinematic cuts applied and the physics motivation
behind them.

\singlespaced

\begin{itemize}
\item[1.] Three isolated leptons (electrons or muons)
          to reduce QCD backgrounds with $p_T$ cuts of
          ($p_T^{(1)}$~;~$p_T^{(2)}$~;~$p_T^{(3)}) > (10~;~5~;~5$)~GeV
          for the highest (1), next highest (2),
          and lowest (3) $p_T$ leptons.
          $p_T^{(1)} >$ 10 GeV is necessary for triggering.
\item[2.] $|m_{\ell^+\ell^-}-m_Z | > 15\gev$ for opposite sign, same
          flavor lepton pair invariant masses
          to reduce $Z\ra e^\pm e^\mp,\mu^\pm\mu^\mp$ backgrounds.
\item[3.] $m_{\ell\ell'} >$ 20 GeV for all lepton pair invariant masses
          to reduce $\gamma^*$, $b \ra \ell+X$.
\item[4.] Opposite sign $\ell^{(1)}$, $\ell^{(2)}$ (highest and next highest
          $p_T$ leptons) must not be ``back to back'', $|\phi_2-\phi_1|<2.5$
          to reduce $Z\ra\tau^\pm\tau^\mp$ background.
\item[5.] Transverse mass required to be $m_T <$ 70 GeV
          to reduce $W$ backgrounds, where $m_T$ is constructed
          from $\ell^{(1)}$'s momentum and the $\slashchar{\vec{E}}_T$
          vector.
\end{itemize}

\doublespaced

Table~\ref{trileptons-table} shows the reduction of
backgrounds as a function
of the cuts 1--5.  In addition, we find no backgrounds from
$W+g$ nor $t\overline{b}+b\overline{t}$, which would require two heavy
quarks to decay to isolated leptons.  This reinforces our confidence
that we need only consider the background from a single heavy quark decay
to an isolated lepton.  After applying all of the cuts
listed above, our final trilepton background estimate is 0.67 fb for
the Fermilab Tevatron $p\overline{p}$ collider at $\sqrt{s}=2\tev$\@.

\begin{table}
\begin{center}
\begin{tabular}{|l|c|c|c|c|r|} \hline
\multicolumn{6}{|c|}{\bf Trilepton Background Results}\\ \hline
        & \multicolumn{5}{c|}{ $\sigma$ after cuts (fb) }\\ \hline
Process & cut 1 & cuts 1-2 & cuts 1-3 & cuts 1-4 & cuts 1-5 \\ \hline\hline
$WZ$, $W\gamma^*$                       & 22.6 & 1.3 & 1.0  & 0.85 & 0.38 \\
$ZZ$, $Z\gamma^*$, $\gamma^*\gamma^*$   &  5.3 &  0.22 & 0.16 & 0.12 & 0.09 \\
$t\overline{t}$                         &  0.42 & 0.33 & 0.23 & 0.17 & 0.06 \\
$Z+g$, $\gamma^*+g$              &  5.0 &  0.64 & 0.23 & 0.14 & 0.14 \\ \hline
Total                  & 33.3 &  2.5 & 1.62 & 1.28 & {\bf 0.67} \\ \hline\hline
\end{tabular}
\caption{Summary of our background studies of the supersymmetric
trilepton signal.  Cuts 1--5 are described in the text.  The final
background estimate using our cuts is 0.67 fb.}
\label{trileptons-table}
\end{center}
\end{table}

Using this background estimate, we can calculate the
smallest SUSY trilepton cross section, folded with the leptonic
branching ratios and detection efficiency (denoted $\sigbreff$),
that is detectable above backgrounds.  We find the minimum $\sigbreff$
for integrated
luminosities of 200 ${\rm pb}^{-1}$, 2 ${\rm fb}^{-1}$
and 25 ${\rm fb}^{-1}$ at the
Fermilab Tevatron to be 25 fb, 3.0 fb
and 0.82 fb respectively, based on the larger of 5 events
or the number of events required for a $5 \sigma$ significance
above background.

\begin{figure}
\centering
\epsfxsize=6in
\hspace*{0in}
\epsffile{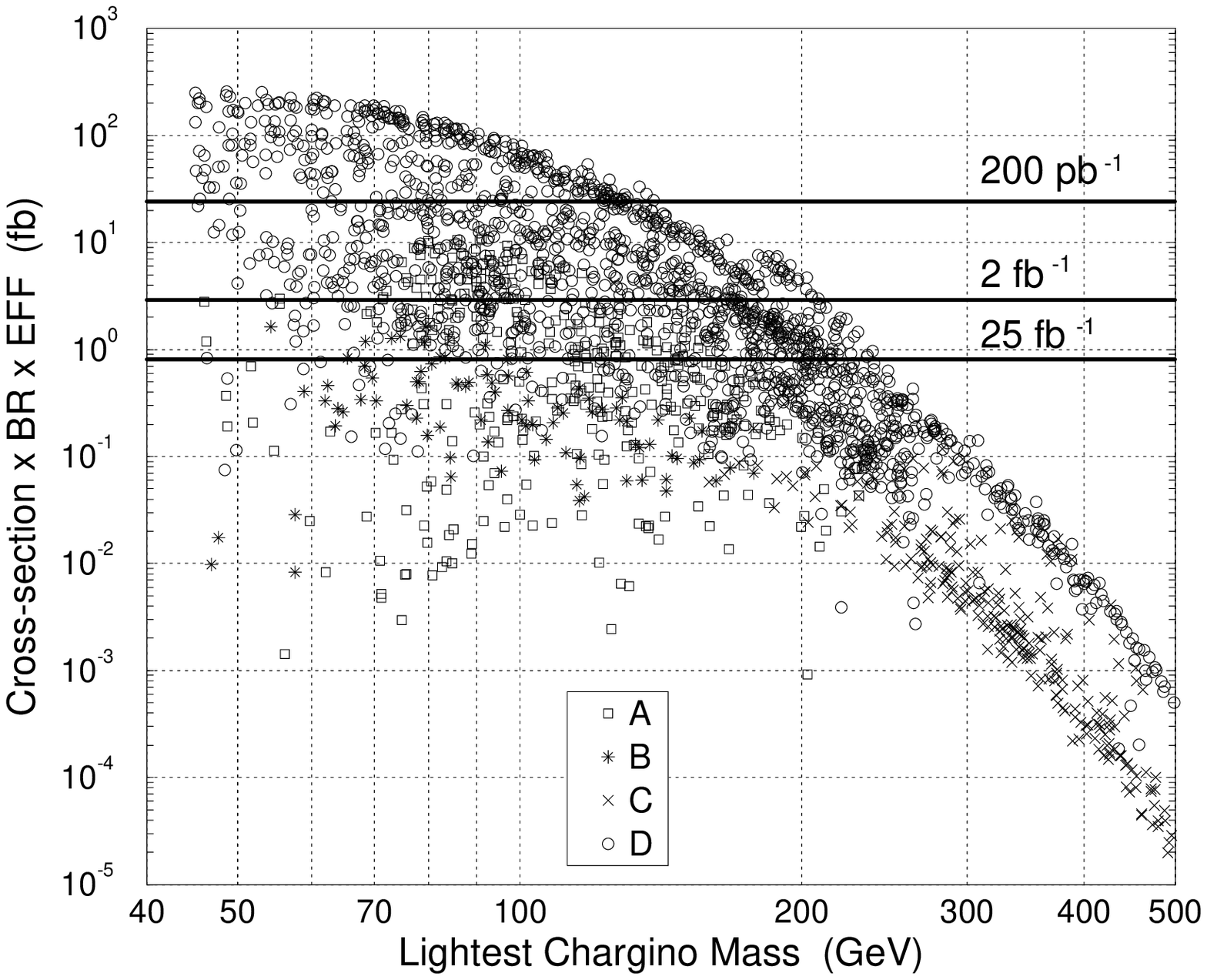}
\caption{Total supersymmetric trilepton signal ($\sigbreff$) after cuts
versus the lightest chargino mass in the CMSSM.  The branching
ratio (BR) is defined as the fraction of $\charI\neutII$ events that decay
to 3 leptons.  The efficiency (EFF) is defined as the fraction
of 3 lepton events that pass the cuts described in the text.  The
minimum detectable $\sigbreff$ for integrated luminosities of
200 ${\rm pb}^{-1}$, 2 ${\rm fb}^{-1}$ and 25 ${\rm fb}^{-1}$ is
shown by the dark horizontal lines at 25 fb, 3.0 fb, and 0.82 fb
respectively.
The different symbols refer to solutions showing
interesting behavior where the second lightest neutralino ($\neutII$) has
(A) a neutral ``invisible'' branching ratio
    (generally $\neutII \ra \snu\nubar$ then $\snu \ra \neutI \nu$) $> 90\%$,
(B) a large destructive interference in 3--body leptonic decays
    defined by $R_{\rm interference} < 0.1$ (see Fig.~\ref{interference}),
(C) a branching ratio to Higgs $> 50\%$ dominates, or
(D) all other solutions.
}
\label{trileptons}
\end{figure}

In Fig.~\ref{trileptons} we have plotted the total
supersymmetric $\sigbreff$ from $\charI\neutII$ production (including
all diagrams) versus the lightest chargino ($\charI$)
mass for well over 2000 solutions spanning the CMSSM parameter space
up to $m_{\charI} \leq 500$ GeV\@.  Note that a tail of CMSSM solutions
{\em does\/} exist for $m_{\charI} > 500$ GeV, but we do not consider
them here.
Each symbol represents one solution defined by $m_0$, $m_{1/2}$, $A_0$,
$\tan{\beta}$ and ${\rm sign}(\mu)$.  Each solution, when the masses and
couplings are run from the unification scale to the weak scale,
has its own well--defined masses, mixings, branching ratios, etc.
The chargino mass represents one of these well--defined weak scale
observables that is directly related to the production cross section
(hence, our choice of x-axis).  In addition, the branching ratios
of the superpartners also represent
weak scale observables that are crucial to a correct calculation of
the trilepton signal.  All decay channels of all possible superpartners
that could be produced directly ($\charI$, $\neutII$), or as a
result of decays (gauginos, sleptons and Higgs), must be computed for
each solution.

\begin{figure}
\centering
\epsfxsize=3in
\hspace*{0in}
\epsffile{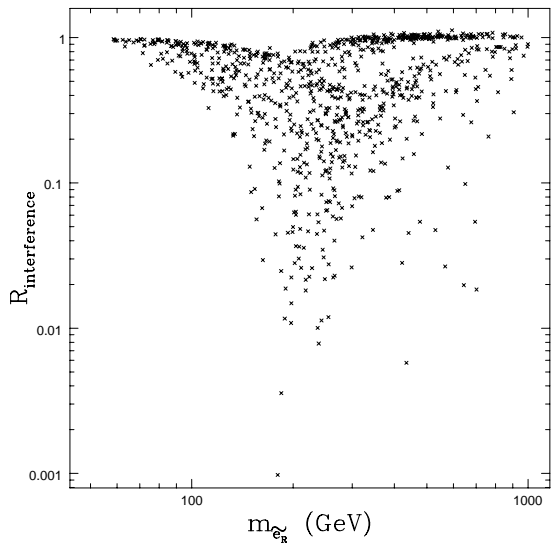}
\caption{To illustrate the effects of interference in the 3--body
decays $\neutII \ra \neutI \ell^+\ell^-$, we have plotted the
ratio ($R_{\rm interference}$) of the full Z, ${ \widetilde e}_R$
and ${ \widetilde e}_L$ squared--amplitude to the sum of the
squares of the three individual diagrams for the branching ratio to
electrons (which is also representative of the branching
ratio to muons and taus) against the mass of the right selectron.
Note that $R_{\rm interference}$ is expected to be $\sim$1 if the
magnitude of the interference is small.
}
\label{interference}
\end{figure}

The second lightest neutralino's leptonic branching ratio can
sometimes be small enough to prevent a detectable trilepton signal
even at low neutralino (or chargino) masses.  Here we elaborate
on the neutralino ($\neutII$) branching ratios while examining in detail
the solutions identified in Fig.~\ref{trileptons}
that have a small leptonic branching ratio.
(A) The $\neutII$ often decays predominantly
to neutral (``invisible'') products through the 2--body
$\snu\nubar$ channel followed by $\snu \to \neutI\nu$,
or directly through the 3--body $\neutI\nu\nubar$.
We have identified these solutions in
Fig.~\ref{trileptons} with separate symbols to clearly show they
are a source of some low $\sigbreff$ solutions.
(B) The 3--body leptonic decays of the $\neutII$
can have destructive interference among the diagrams mediated by the
$\lL$, $\lR$ and the $Z$.  This interference effect has also been
observed in Ref.~\cite{baer95:9504234}.  Once again, we identify such
solutions in Fig.~\ref{trileptons} with a separate symbol.
Note that significant destructive interference occurs only for chargino
masses $m_{\charI} \lsim 200$ GeV\@.  Generally, the interference
is characterized by $m_{\lL}, m_{\lR} \sim 200$ GeV\@.
At lower slepton masses, it is the slepton mediated 3--body decays
that dominate the decay width of $\neutII$ into leptons.  At higher
slepton masses, the $Z$ mediated 3--body decays dominate.
This is demonstrated in
Fig.~\ref{interference}, where the ratio of the squared--amplitude
with and without the interference terms for
$\neutII \ra \neutI e^+ e^-$ (which is representative of
all the leptonic channels) is plotted versus the right
selectron mass for all solutions described above.
We find destructive interference for {\em both\/} signs
of $\mu$.  The interference for $\mu > 0$ is largest for
$\tan \beta \gsim 10$, while the interference for $\mu < 0$ occurs over
the entire range of $\tan \beta$.
Finally, (C) the decays of $\neutII \ra \neutI \h$ have been explicitly
identified in Fig.~\ref{trileptons}.
We note that to compute the decay widths of $\h$,
one must evaluate all masses (and couplings) at the scale of
$m_{\h}$~\cite{braaten80:715}.
Since there are completely general upper mass bounds~\cite{kane93:2686}
on $m_{\h} \ll 2 m_t$, the only decay channels open to $\h$ are to
$b\overline{b}$, $\tau^+\tau^-$ and lighter particles.
When $m_b$ and $m_{\tau}$ are evaluated
at $m_{\h}$, the leptonic decay width of $\h \ra \tau^+\tau^-$
is about 10\%.
This has a dramatic effect on the leptonic branching ratio of the
$\neutII$ when both $\tau$'s decay leptonically to $e$'s and $\mu$'s.
Further, it is likely that 1--prong pion decays from the Higgs
are measurable and could be added to the branching
fraction; thus all the $\times$'s in Fig.~\ref{trileptons}
(which have a large branching ratio to $\h$) would move up
in cross section by a factor of about 2--3.
However, accurate estimates of 1--prong backgrounds
are more difficult than leptonic backgrounds, so to be conservative
we do not include isolated pions in our signal.
The final result is that we find the
trilepton signal is present and detectable given a high enough
integrated luminosity even for large neutralino masses
($m_{\neutII} \sim m_{\charI} \gsim 250$ GeV) where the
$\neutII \ra \neutI \h$ decay channel is open.

\begin{figure}
\centering
\epsfxsize=6in
\hspace*{0in}
\epsffile{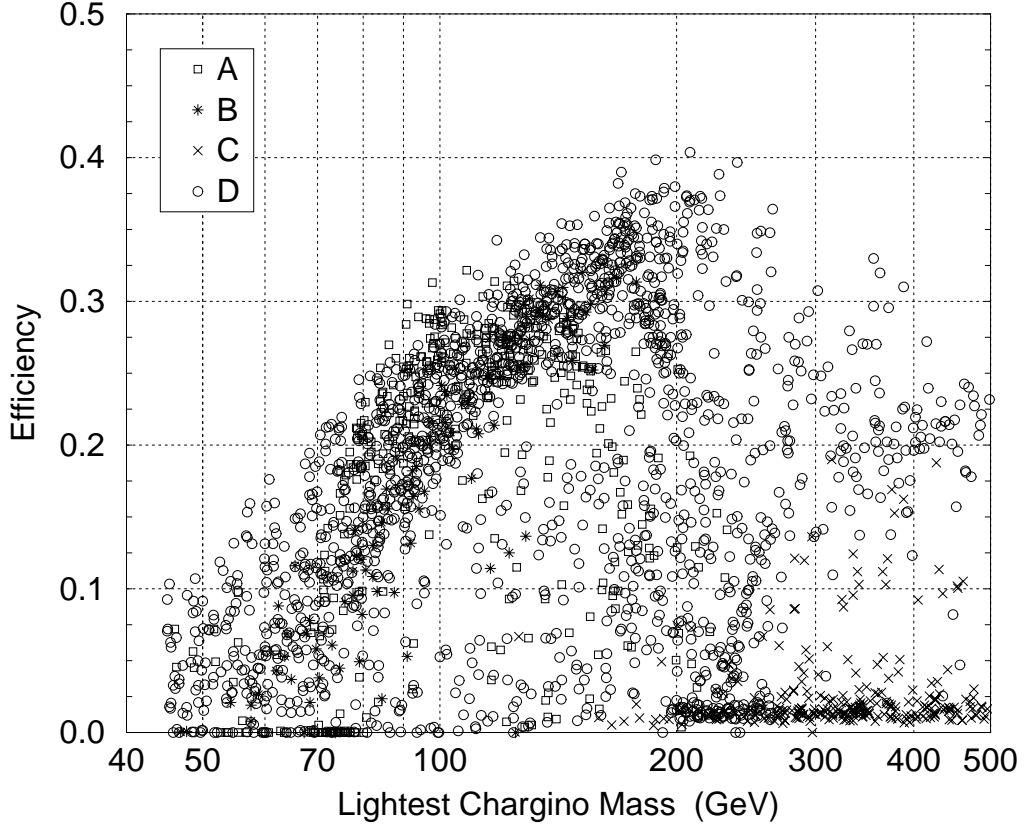}
\caption{Detection efficiency of the trilepton signal after cuts
versus the lightest chargino mass in the CMSSM.  The efficiency
is defined as the fraction of 3 lepton events that pass the cuts
described in the text.
The different symbols refer to solutions showing
interesting behavior where the second lightest neutralino ($\neutII$) has
(A) a neutral ``invisible'' branching ratio
    (generally $\neutII \ra \snu\nubar$ then $\snu \ra \neutI \nu$) $> 90\%$,
(B) a large destructive interference in 3--body leptonic decays
    defined by $R_{\rm interference} < 0.1$ (see Fig.~\ref{interference}),
(C) a branching ratio to Higgs $> 50\%$ dominates, or
(D) all other solutions.
}
\label{efficiency}
\end{figure}

Once the cross section and branching ratios are computed,
a full event--level simulation was performed using the cuts
described above.  The cuts undoubtedly have an impact on
the detectable signal, and we quantify this by defining the
``detection efficiency'' as the
ratio of the number of 3 lepton events that passed our cuts to
the total number of 3 lepton events from $\charI\neutII$
production.  We find that the efficiency can vary dramatically
over the parameter space, as is illustrated in Fig.~\ref{efficiency}
where we plot the efficiency versus the chargino mass.
The structure of the efficiency plot is clear:  At low chargino
masses ($m_{\charI} \lsim 150$ GeV) most of the $\charI$, $\neutII$
decays are 3--body with a gradual rise in efficiency due to
the increasing energy of the leptons.  The low efficiency
solutions (EFF $\lsim 10^{-2}$, where we found the EFF to be $> 10^{-4}$
for {\em all\/} solutions) occur when
$m_{\charI} \approx m_{\snu}$ and the 2--body sneutrino modes
are open and there is little energy for the lepton.
The neutralino also has analogous problems
when $m_{\neutII} \approx m_{\lL \, {\rm or} \, \lR}$ so that
2--body modes are open with one rather soft lepton.
At higher chargino masses ($m_{\charI} \gsim 200$ GeV),
the 2--body decays
$\charI \ra \neutI W$ and $\neutII \ra \neutI Z$ also cause a lower
efficiency for the simple reason that our {\em cuts\/} are
designed to eliminate real $W$'s and $Z$'s.  Finally, the
somewhat lower efficiency of the $\neutII \ra \neutI \h$ mode
is visible at high chargino masses ($m_{\charI} \gsim 250$ GeV)\@.
Note that the efficiencies described here
(visible in Fig.~\ref{efficiency}) are automatically
included in the $\sigbreff$ plot of Fig.~\ref{trileptons}.

The results from this trilepton analysis are manifest in
Fig.~\ref{trileptons}.  The Fermilab Tevatron can probe
chargino masses up to 140 GeV, 210 GeV and
240 GeV with integrated luminosities of 200 ${\rm pb}^{-1}$,
2 ${\rm fb}^{-1}$ and 25 ${\rm fb}^{-1}$ respectively.  However,
the Tevatron {\em cannot\/} set mass limits on charginos
or neutralinos from {\em just\/} the trilepton signal.
This result follows directly from
using the full CMSSM parameter space with the complete
2--body and 3--body branching ratios of superpartners and
a full event--level simulation with realistic detector cuts.
The requirement for a full simulation of all constrained
solutions so that the efficiency is calculated correctly for
every solution is demonstrated in Fig.~\ref{efficiency}.
Perhaps new experimental or theoretical constraints
could eventually eliminate the low lying solutions in
Fig.~\ref{trileptons}, thereby allowing a lower limit
on $m_{\charI}$ to be set if no signal is found.
Nevertheless, we emphasize that a
significant part of the parameter space can be eliminated
if no signal is found.

\subsection{Effects of Constraints on Supersymmetric Parameter Space}
\indent

Requiring a constrained parameter space that
satisfies our theoretical expectations with all current
experimental results is very important.  To illustrate
the effect of constraints on the parameter space, we looked at
a particular choice of parameters that would give
a qualitative feel for how $b\to s\gamma$ and relic density
cuts in the CMSSM typically impact the trilepton signal.
For this example, $\tan\beta=5$, $A_0=0$, and ${\rm sign}(\mu)=+$
(only for illustrative purposes in this section), while
the values of $m_0$ and $m_{1/2}$ were selected randomly
on a logarithmic scale up to $1\tev$.  We applied all CMSSM
cuts to the data sample except the $b\to s\gamma$ and
relic density cuts.

\begin{figure}
\centering
\epsfxsize=5in
\hspace*{0in}
\epsffile{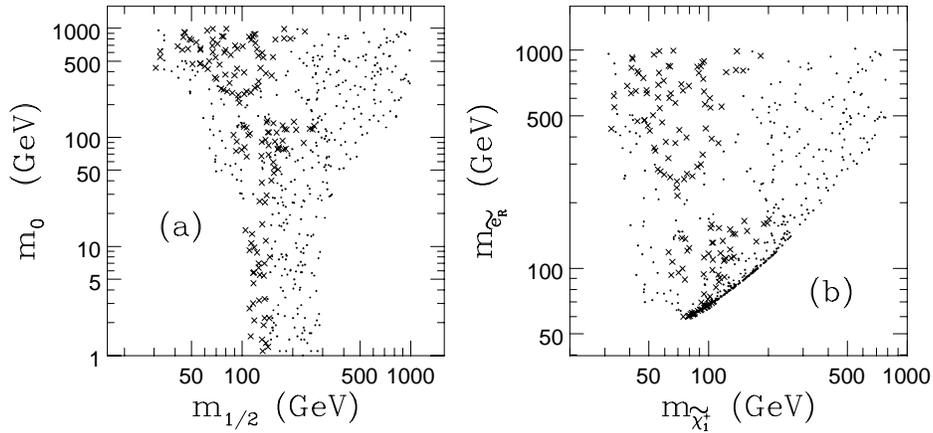}
\caption{The trilepton signal before $b\to s\gamma$ and relic density cuts.
The solutions represented by the $\times$'s are detectable with an integrated
luminosity of 25 ${\rm fb}^{-1}$ at the Fermilab Tevatron, while the
dots are not.}
\label{tri_all}
\end{figure}

In Figs.~\ref{tri_all}a and~\ref{tri_all}b
we plot the trilepton signal {\em before\/} the $b\to s\gamma$
and relic density cuts.
The plotting variables are $m_0$ (common scalar mass)
and $m_{1/2}$ (common gaugino mass)
in Fig.~\ref{tri_all}a and the more concrete $m_{\tilde e_R}$
and $m_{\charI}$ in Fig.~\ref{tri_all}b.
(Fig.~\ref{tri_all}b is nothing more than a direct re-mapping
of all the points in Fig.~\ref{tri_all}a.)  The $\times$'s in
Figs.~\ref{tri_all}a
and~\ref{tri_all}b are solutions which are detectable
through the supersymmetric
trilepton signal with an integrated luminosity of 25 ${\rm fb}^{-1}$,
while the dots are not detectable without higher integrated luminosity.

\begin{figure}
\centering
\epsfxsize=5in
\hspace*{0in}
\epsffile{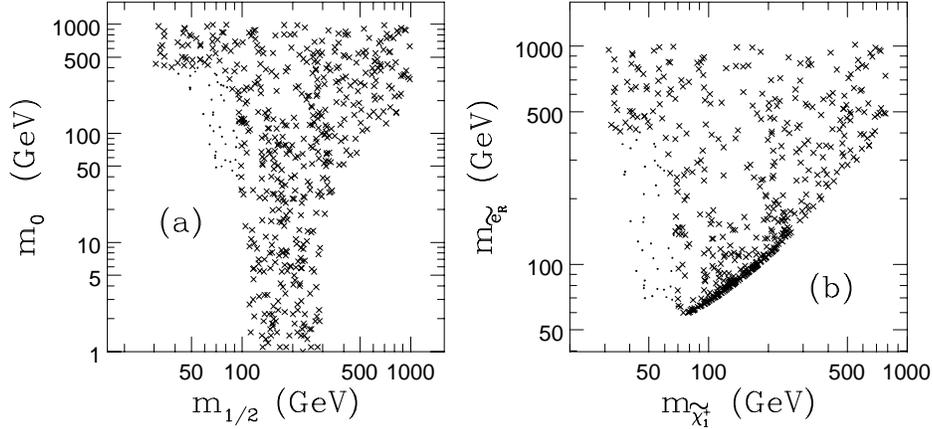}
\caption{The effect of $b\to s\gamma$ cut on the parameter space.
The solutions represented by the $\times$'s satisfy the cut
Br($b \ra s\gamma$) $< 5.4 \times 10^{-4}$, while
the dots do not.}
\label{bsg}
\end{figure}

In Figs.~\ref{bsg}a and~\ref{bsg}b we produce the same type of scatter
plots, this time using the $b\to s\gamma$ observable.
All solutions marked with an $\times$ are now {\em just\/} those with
the additional constraint ${\rm Br}(b\to s\gamma ) < 5.4\times 10^{-4}$.
We choose $5.4\times 10^{-4}$ as our discriminant value since that is
the $95\%$ upper limit value published by CLEO~\cite{cleo93:674}.
CLEO has also recently
reported~\cite{cleo95:2885} a
measurement of
\beq
{\rm Br}(b\to s\gamma)=(2.32\pm 0.67)\times 10^{-4}.
\eeq
This would suggest that
we could add $0.67\times 10^{-4}$
to the central value and declare this our $1\sigma$ upper bound on
${\rm Br}(b\to s\gamma)$, and then insist that all CMSSM solutions
have a predicted value of ${\rm Br}(b\to s\gamma)<2.99\times 10^{-4}$.
However, the QCD uncertainties~\cite{buras94:374}
in the ${\rm Br}(b\to s\gamma)$ calculation (approximately $25\%$)
require us to use
a higher ``calculation upper limit'' than the experimental upper limit;
therefore, to be conservative we accept all solutions with a {\em calculated\/}
${\rm Br}(b\to s\gamma) < 5.4\times 10^{-4}$.  A significant number of
solutions in Figs.~\ref{bsg}a and~\ref{bsg}b are cut out by this
constraint.  Interestingly, many of the solutions cut out by this
constraint overlap with light chargino solutions
which were {\em not detectable\/} by the trilepton signal.

The relic density constraint is also an important cut on the CMSSM parameter
space.  We expect that nature is described by an R--parity
conserving supersymmetric theory~\cite{martin92:2769,diehl95:399},
and thus the lightest
supersymmetric particle (LSP) is absolutely stable.  In the early universe
these stable particles were in thermal equilibrium with the photons
until the expansion rate of the universe became roughly
equal to their annihilation rate.  When this happens, the LSPs fall out
of equilibrium with the photons and their relic abundance stabilizes.
If this decoupling occurs too soon (weak annihilation rates) then the
universe becomes matter dominated too early.
Most observational data indicates that the universe is more
than about 10 billion years old, and this translates into an upper bound
on the relic density of LSPs.  Quantitatively this upper bound, often
called the ``age of the universe constraint'', is most
often expressed as a condition on $\Omega_{LSP} h^2$:
\beq
\Omega_{LSP} h^2<1.0~~~~{\rm (age~of~the~universe~constraint)}
\eeq
where $h$ is the Hubble parameter.  As part of the CMSSM we require
$\Omega_{LSP} h^2 < 1.0$.  Except for a few possible pathological
cases~\cite{drees93:376,leszek94:susy94}, the age of the universe
constraint will
exclude all SUSY models with large squark and slepton
masses~\cite{drees93:376,kane94:6173}.
This can be understood by realizing that the LSP is mostly
(but not completely) bino
in the CMSSM, and the bino has no coupling to the $Z$ boson and
therefore no annihilation channels through an $s$--channel $Z$ are
accessible.  Then the LSP must annihilate through a $t$--channel
squark or slepton:
\beq
\langle \sigma v_{rel}\rangle \propto \frac{1}{m^4_{\tilde f}}.
\eeq
And since
\beq
\Omega_{LSP} h^2 \propto \frac{1}{\langle \sigma v_{rel} \rangle}~~~~~~
{\rm then}~~~~~~\Omega_{LSP} h^2 \propto m^4_{\tilde f}.
\eeq
Therefore, the relic abundance $\Omega_{LSP}$ grows with the supersymmetric
breaking scale.  This cutoff on supersymmetric masses depends on all the
input parameters but is generally around $1\tev$.

\begin{figure}
\centering
\epsfxsize=5in
\hspace*{0in}
\epsffile{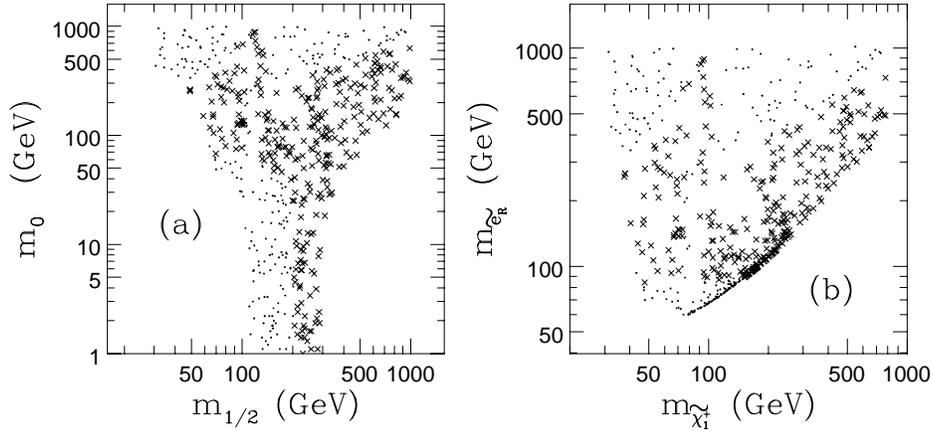}
\caption{The effect of the $\Omega_{LSP} h^2>0.05$ cold dark matter cut
and $\Omega_{LSP} h^2<1.0$ age of the universe constraint
cuts on the parameter space.  The solutions represented by
$\times$'s have $0.05 < \Omega_{LSP} h^2 < 1.0$ while the dots do not.
}
\label{relic}
\end{figure}

The importance of this robust
cosmological requirement on studies of supersymmetry
detectability cannot be overstated.  Any successful constraint on the
mass of supersymmetric particles is obviously of great
relevance to a collider program trying to discover or rule out
supersymmetry.  It is for this reason that we have ensured that
this constraint is incorporated into all solutions analyzed
in this paper.

\begin{figure}
\centering
\epsfxsize=5in
\hspace*{0in}
\epsffile{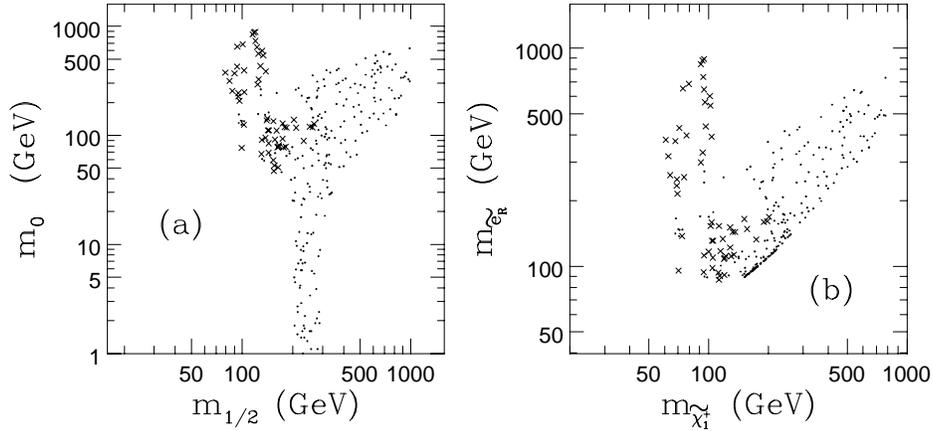}
\caption{The trilepton signal is plotted as in Fig.~\ref{tri_all}
{\em except\/} all solutions ($\times$'s and dots) must
satisfy the $b\to s\gamma$ cut and relic density cuts described
in the text.  Note the large regions of parameter space excluded
by these cuts, and also the remaining distribution of
detectable solutions (represented by $\times$'s for an integrated
luminosity of 25 ${\rm fb}^{-1}$, as in Fig.~\protect\ref{tri_all}).
}
\label{tri_cut}
\end{figure}

A lower bound on $\Omega_{LSP} h^2$ can also
be obtained from the requirement that the LSPs constitute a significant
amount of the universe's mass fraction to be a viable cold dark matter
candidate.  Using this requirement, we obtain a quantitative lower bound,
\beq
\Omega_{LSP} h^2 \gsim 0.05~~~~~~{\rm (cold~dark~matter~constraint).}
\eeq
This cold dark matter constraint is not applied to the CMSSM solutions
anywhere else in the paper except in Figs.~\ref{relic} and~\ref{tri_cut},
since no experiment has yet confirmed the identity of the
cold dark matter.  However, we do consider it a major success
of the unified supersymmetric theories that
a stable weakly interacting massive object with a large
relic abundance is predicted in accordance with astrophysical
observations.

In Figs.~\ref{relic}a and~\ref{relic}b the $\times$'s represent
solutions which are consistent with both the age of the universe and
the cold dark matter constraints, summarized by
$0.05< \Omega_{LSP} h^2<1.0$.
The $\times$'s represent all solutions which
pass the relic density cut, and the dots represent all solutions
which lie outside the cut.
The dots in the lower left corners of both
Figs.~\ref{relic}a and~\ref{relic}b are solutions with
$\Omega_{LSP} h^2 < 0.05$, and thus are not interesting dark matter
candidates.  The dots with high $m_0$ and high $m_{\tilde e_R}$ are
solutions which have $\Omega_{LSP} h^2$ in qualitative agreement
with the argument given above that large SUSY scalar masses yield
large $\Omega_{LSP} h^2$.  Some solutions at high $m_0$ survive
at $m_{1/2}\sim 100\gev$ where the LSPs annihilate through
the $Z$ resonance.  Although the coupling to the $Z$ is quite small, the
resonance effect is dominant here and the relic density can stay small.

Finally, Figs.~\ref{tri_cut}a and~\ref{tri_cut}b are identical to
Figs.~\ref{tri_all}a and~\ref{tri_all}b except now
the ${\rm Br}(b\to s\gamma)$
and $\Omega_{LSP} h^2$ cuts are included.
Many more models are ruled
out from the additional cuts, and the resulting parameter space
provides a more constrained set of solutions
that the Fermilab Tevatron could either detect or
rule out (represented by the $\times$'s).  The qualitative shift
in the parameter space
by including all constraints on the CMSSM demonstrated in this
example points to the importance of including all relevant
constraints simultaneously on supersymmetric solutions for a
realistic study of collider capabilities.

\subsection{Missing $E_T$ + Jets as a Signal for Squarks and Gluinos}

\indent
A classic signature of supersymmetry is multi--jet events with a large
$\Et$~\cite{missingetjets,baer92:142,baer95:2159}.
This signal can result from many supersymmetric parton--level processes, and
we have simulated only a practical subset of these:
squark/gluino pair (${\tilde q}{\tilde q}$,
${\tilde g}{\tilde g}$), squark+gluino (${\tilde q}{\tilde g}$), and
squark/gluino + chargino/neutralino production
(${\tilde q}{\tilde \chi}$, ${\tilde g}{\tilde \chi}$).
In the special case of $m_{1/2} \gg m_0$ then
$m_{\tilde g} \approx m_{\tilde q}$, otherwise
$m_{\tilde g} \lsim m_{\tilde q}$ and we expect
\begin{figure}
\centering
\epsfxsize=3in
\hspace*{0in}
\epsffile{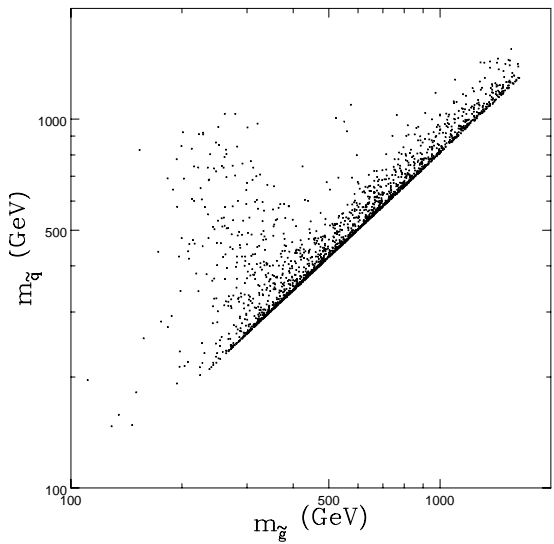}
\caption{The correlation between the squark ($\tilde q$) and gluino
($\tilde g$) mass is shown for all CMSSM solutions.  We have plotted
$\tilde q = \tilde u_L$ to be representative of all the squarks.  Note
that relatively few solutions have $m_{\tilde q} = m_{\tilde g}$
or $m_{\tilde q} \gg m_{\tilde g}$.
}
\label{gluino-squark}
\end{figure}
the dominant signal to be $\tilde g \tilde g$ production with
3--body decays into other gauginos and jets in that region
(see Fig.~\ref{gluino-squark}).
It is expected that the background is dominated by QCD multi--jet
production folded with the intrinsic $\Et$ resolution of the detector.
To estimate the background contribution, we generate all
QCD parton--level processes for $\hat{p}_T > 30$ GeV with
initial and final state QCD radiation.
The $\Et$ resolution $\sigma_{\Et}$ is
approximated by the CDF--like formula
$\sigma_{\Et} = 0.7\sqrt{\Sigma E_T}$,
where $\Sigma E_T$ is the scalar sum of all the jet transverse energy.
The $\Et$ resolution degrades with increased jet activity,
so only high $p_T$ events can generate a large $\slashchar{E}_{T}$.
The contribution of each event is folded with the probability that
the measured $\Et$ fluctuates to $\Etcut$ or more
using the Gaussian formula:
\beq
P(\Et > \Etcut) = \frac{2}{\sqrt{\pi}} \int_{t'}^{\infty} e^{-t^2}dt~~~~~~
{\rm where}~~~~~~t' = \frac{\Etcut}{\sigma_{\Et}/\sqrt{2}}.
\eeq
The resultant weighted cross section for
$\Etcut$ = (0,50,75,100,125) GeV is
(0.13 $\mu$b, 1.3 pb, 2 fb,$\,\simeq 0$,$\,\simeq 0$).
However, a Gaussian approximation to the resolution is not realistic, since
non--Gaussian tails are known to be important.  To test the sensitivity of
our estimate, we have added a term to our probability distribution
so that the total probability of mismeasurement from 2.58--6.00$\sigma$
is still $\simeq$ 1\% but the probability is evenly distributed.
For this case, our background estimate is (0.13 $\mu$b, 115 pb, 1.8 pb, 103 fb,
$\simeq$ 0).   Because of the strong sensitivity to the non--Gaussian tail,
we choose $\Etcut$ = 75 GeV\@.
This reduction of the background is conservative since we have not
utilized the altered kinematics resulting from the loss of jet energy.
We further apply a transverse sphericity cut $S_T > 0.2$ to reduce this
pure jet background to a negligible level.
The other physics backgrounds we considered all have $\Et$ from
$Z(\ra\nu\bar{\nu})+g$, $Z(\ra\tau^+\tau^-)$, $W(\ra\ell\nu)+g$, and
$t\bar{t}$.
Since we veto events containing isolated muons and electrons with
$p_T > 15$ GeV, $W(\ra\tau\nu)$ is a large potential background.
Backgrounds are also reduced by requiring that the sum
$E_T^{j_1}+E_T^{j_2}+\slashchar{E}_T > 300$ GeV, where
$j_1$ and $j_2$ are the two highest $E_T$ jets.
Finally, we require that $\Delta\phi > 0.5$ between each jet and the
$\Et$ direction to reduce the fake $\Et$
\begin{figure}
\centering
\epsfxsize=4in
\hspace*{0in}
\epsffile{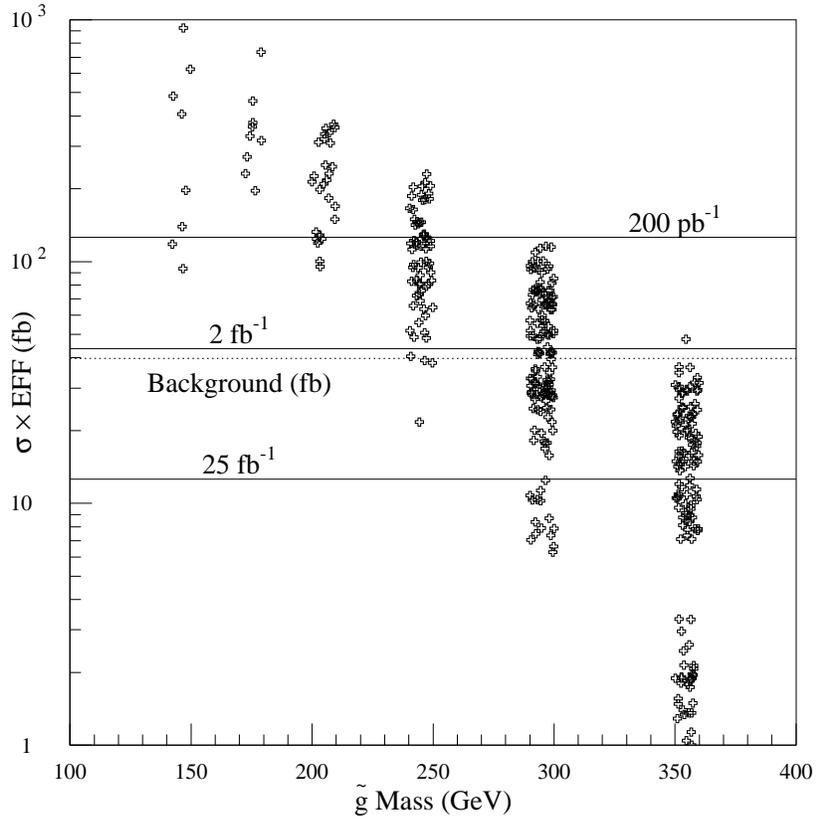}
\caption{The $\sigma$ $\times$ EFF is plotted for the $\Et$ + jets
signal versus the gluino mass.  The efficiency (EFF) is defined
as the fraction of events that pass the cuts described in the text.
Horizontal lines indicate the gluino mass reach
of the Fermilab Tevatron at integrated luminosities of
200 ${\rm pb}^{-1}$, 2 ${\rm fb}^{-1}$, and 25 ${\rm fb}^{-1}$.
The dotted line represents the background cross section with our cuts.
The vertical banding in gluino mass is due to numerical
sampling and is not physically significant.
}
\label{gluino_sigma}
\end{figure}
from energy lost in cracks or shower fluctuations.  Using these cuts,
the cross sections are 5 fb, 11 fb, and 24 fb for the $Z$, $W$,
and $t\bar{t}$ backgrounds respectively.

After a full simulation we find the
gluino mass reach for integrated luminosities of
200 ${\rm pb}^{-1}$, 2 ${\rm fb}^{-1}$ and 25 ${\rm fb}^{-1}$ to be
$\sim$300 GeV, $\sim$350 GeV, and $\sim$400 GeV respectively, at
$10\sigma$ signal significance (see Fig.~\ref{gluino_sigma}).
Here significance is defined as the number of signal events
divided by the square root of the number of background events, for
events that pass the above cuts with $\Etcut = 75$ GeV\@.
Of course, this is not a useful procedure unless
one has reasonable knowledge of the normalization of the experimental
background $\Et$ distribution as well as all the real
backgrounds, which we assume will occur.  Note that we do not make explicit
use of the signal or background shape, just the number of events above
$\Etcut$.

\begin{figure}
\centering
\epsfxsize=2.7in
\hspace*{0in}
\epsffile{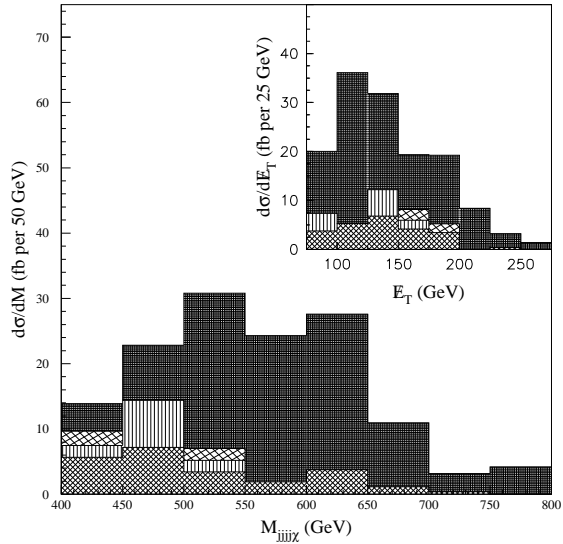}
\caption{The invariant mass distribution ($M_{jjjj\chi}$) found by
summing the four vectors of the 4 highest $E_T$ jets with the four
vector ($\Et$,$\slashchar{ \vec{E} }_T$) is plotted for an
{\em example\/} model with $m_{\tilde g} = 298$ GeV
and $m_{\tilde q} = 311$ GeV.
The dark shaded region on the top is the signal, while the
lower three shaded regions are the backgrounds from $Z$,
$W$ and $t\overline{t}$ production respectively.  The inset figure
shows the corresponding $\Et$ distribution for the same model.
}
\label{gluino}
\end{figure}

We have studied several kinematic variables to see if the
resulting distributions can make the signal more convincing
or help us extract any information about the superpartner masses.
As an example, we choose a model with
$m_{\tilde g} = 298$ GeV and $m_{\tilde q} = 311$ GeV
that has a total production cross section of 1.4 pb.
The invariant mass found by summing the four vectors of the
4 highest $E_T$ jets with the four vector
($\Et$,$\slashchar{ \vec{E} }_T$) is shown in
Fig.~\ref{gluino} for the signal (dark upper bins) and the backgrounds
(lower light or hashed bins).  The inset figure shows the $\Et$
distribution for the same processes.  In the invariant mass spectrum
the signal clearly peaks around
$2 m_{\tilde g} \simeq 2 m_{\tilde q} \sim 600$ GeV,
while there is no such peak in the $\slashchar{E}_T$ distribution.
Since the backgrounds from $Z$, $W$, and $t\overline{t}$
peak at lower invariant masses, one can see how the invariant
mass distribution provides an effective means to separate the
signal from background.

\subsection{Signals of Stop Production}
\indent

The signal from stop production has been known for some time
to be quite promising~\cite{ellis83:248,baer89:303,
baer91:725,baer94:4517,lopez94:9406254}
since the lightest stop mass eigenstate ($\tilde t_1$) is generally lighter
than all the other squarks.  This follows from the substantial mixing
between the top squark weak eigenstates $\tilde t_L$ and $\tilde t_R$
caused by a large mixing term in the stop mass matrix proportional
to the top mass.  Furthermore, the large top Yukawa coupling
in the running of the stop scalar mass
terms reduces the mass of the stop with respect to the other
squark masses.  The result is a large mass splitting between the
stop mass eigenstates $\tilde t_1$ and $\tilde t_2$.
Hence, $\tilde t_1 \tilde t_1^*$
is more readily pair produced than the other squarks.
The decay $\tilde t_1 \ra \charIplus b$ dominates if it is kinematically
allowed, yielding a final state analogous to $t\ra W^+b$ in the SM.
For those models of the CMSSM where this decay is not allowed,
we assume the decay $\tilde t_1 \ra \neutI c$ with a 100\%
branching ratio.

A stop search is similar to a top search if
$\tilde t_1 \ra \charIplus b$, though we take advantage of
the different kinematics to separate stop from the top {\em background\/}.
The full set of backgrounds considered is $t\bar{t}$, $W(\ra\ell\nu)+g$,
$Z(\ra\nu\bar{\nu})+g$, $W^\pm W^\mp$, and $Z(\ra\tau^+\tau^-)$.  Additional
jet activity is generated by gluon splitting and initial and final state
radiation.  We classify the events into 3
channels: 1) dilepton, 2) $W$ + jets, and 3) dijet.
Channels 1) and 2) are distinct from a top signal because
the lepton ($W$ + jets) or leptons (dilepton) have smaller
$p_T$ and an $\Et$ inconsistent with $W$ decay.
Dijet events 3) are $\Et$ + jets events with no isolated leptons.
We identify dilepton events by the following kinematic cuts:

\singlespaced
\begin{itemize}
\item[1.] Two isolated electrons or muons with $E_T > 10$ GeV\@.
\item[2.] Two or more jets, one or two $b$--tagged.
\item[3.] $\Et > 25$ GeV; if $25 < \Et < 50$ GeV then an additional
          cut $\Delta\phi > 0.3$ is applied, where
          $\Delta\phi$ is measured between
          $\slashchar{\vec{E}}_T$ and any of the jets.
\item[4.] $|m_{\ell\ell^{'}} - m_Z | > 15$ GeV,
          and $m_{\ell\ell^{'}}>$ 10 GeV for
          leptons $\ell,\ell^{'}$.
\item[5.] $m_{\ell,b}^{\rm MAX} < 100$ GeV,
          where $m_{\ell,b}^{\rm MAX}$ is the highest invariant
          mass between the leptons and the $b$--jets.
\end{itemize}
$W$ + jets events are defined by:
\begin{itemize}
\item[1.] One isolated electron or muon with $E_T > 10$ GeV\@.
\item[2.] Two or more jets, one or two $b$--tagged.
\item[3.] $\Et > 25$ GeV\@.
\item[4.] $m_T < 75$ GeV, where the transverse mass $m_T$ is constructed from
          the lepton momentum and the $\slashchar{\vec{E}}_T$.
\item[5.] $m_{\ell,b}^{\rm MAX} < 100$ GeV,
          where $m_{\ell,b}^{\rm MAX}$ is the highest invariant
          mass between the leptons and the $b$--jets.
\end{itemize}
Finally, dijet events are classified by:
\begin{itemize}
\item[1.] Two or three jets with $E_T^{j}>$ 20 GeV, and no isolated leptons.
\item[2.] $\Et > 75$ GeV\@.
\item[3.] $\Delta\phi_{j_1 j_2} <$ 3.
\item[4.] $\Sigma E_T^{j}<$ 150 GeV\@.
\item[5.] $m_{j_1,j_2} > 120$ GeV, where $m_{j_1,j_2}$ is the
          invariant mass of the highest ($j_1$) and next highest ($j_2$)
          $E_T$ jets.
\end{itemize}
\doublespaced

\begin{figure}
\centering
\epsfxsize=4.5in
\hspace*{0in}
\epsffile{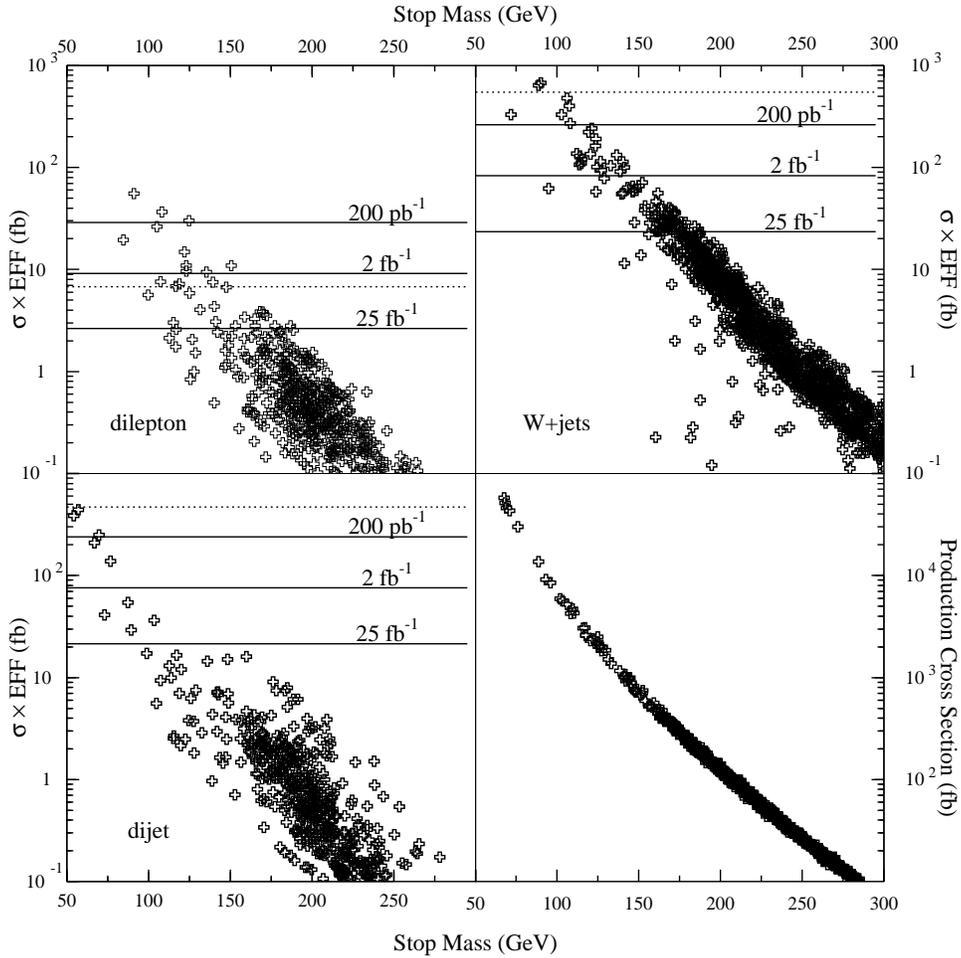}
\caption{The $\sigma$ $\times$ EFF for stop
production is plotted versus the stop mass in the dilepton,
W+jets and dijet channels.
The efficiency (EFF) is defined as the fraction of events
that pass the cuts described in the text for each channel.
Horizontal lines indicate the
reach of the Fermilab Tevatron at integrated luminosities of
200 ${\rm pb}^{-1}$, 2 ${\rm fb}^{-1}$, and 25 ${\rm fb}^{-1}$.
The dotted lines represent the total background cross section for
each channel with our cuts.  Note the total stop production
cross section (with no cuts) is also shown for comparison.
}
\label{stopfig1}
\end{figure}
For event types 1) and 2), where a $b$--tag is required, we assume an
efficiency $\epsilon_b$ {\em per single tag\/} independent of $p_T$ or
$\eta$ of the tracks.
Using $\epsilon_b = 0.3$, the probability to tag one or two $b$'s is then
roughly 0.5.  For dijet events, it was shown previously that
soft lepton tagging of the $c$--jet does not dramatically improve the
stop search \cite{baer94:4517}, so we do not consider it here.

Our results are illustrated in Fig.~\ref{stopfig1}, where the
cross section folded with the detection efficiency
($\sigma \times {\rm EFF}$) is plotted versus the stop mass
for the dilepton, $W$ + jets, and dijet channels.
We find that for an integrated luminosity of 2 ${\rm fb}^{-1}$,
the stop mass reach is up to $\sim$160 GeV at $5\sigma$
significance in the $W$ + jets channel.
For 25 ${\rm fb}^{-1}$, the stop mass reach is up to
$\sim$200 GeV at $5\sigma$ significance in the $W$ + jets channel
and the dilepton channel.  The few models with
$\tilde t_1 \ra \neutI c$ are visible in the upper left
corner of the dijet channel graph.

\subsection{Comment on Top Decays to Stop Signals}
\indent

Another interesting signal of top squarks is through the decay of
the top quark into a top squark and an LSP
($t\to \tilde t_1 \neutI$)~\cite{ellis83:248,hidaka92:155}.
We have analyzed all of the CMSSM solutions to find those
which kinematically allow this decay, and which
also change the branching fraction of $t \ra W^+ b$ significantly
(however, see Section 4).
All solutions which satisfy the kinematic requirement also
give ${\rm Br}(t \ra \tilde t_1 \neutI )\lsim 15\%$.  Since the
$\neutI$ is invisible and the stop will be hard to
see with the soft jets originating through
$\tilde t_1 \ra \neutI c$ decays, perhaps the only
way to detect the effects of the light stops and LSPs
is through modification of the ${\rm Br}(t \ra W^+ b)\simeq 1$
branching ratio.

The best way to test for the existence of these stops
is through stop production.  Since in the
$m_t > m_{\tilde t_1} + m_{\neutI}$ region of
parameter space the stops will decay primarily through a one--loop
diagram to a charm and an LSP, good charm tagging would be extremely
helpful if this signal has a chance of being distinguished from
other copious jet sources.

\section{Light Charginos and Stops Beyond the CMSSM}
\indent

All analyses in this paper are based on the constrained
minimal supersymmetric Standard Model (CMSSM) described
in the introduction.  The CMSSM is more general than the
SM since it includes the SM fully, it is a consistent theory,
it incorporates phenomena not explained by the SM such as
the apparent unification of the gauge couplings, and provides
a derivation of the Higgs mechanism rather than assuming it
as in the SM\@.  Further, we have seen from Section 3 that the
CMSSM makes a variety of testable predictions.

However, there is a hint that the CMSSM may not be entirely correct
because its theoretical assumptions about supersymmetric parameters
are too restrictive.  The purpose of this section is to point out
this possibility and the implications for supersymmetry searches
at the Fermilab Tevatron.

There are two effects that are clues to go beyond the CMSSM\@.  First, the
Br($Z \ra b{\overline b}$) is larger than its SM value by about
$2\sigma$.  Second, the value of $\alpha_s$ deduced in LEP analyses
from the $Z$ width is larger by about $2\sigma$ than the value
of $\alpha_s$ deduced other ways.  While neither of these effects
is of great ``statistical'' significance, both are based on a number
of independent measurements over several years from several detectors;
the errors are mostly systematic and theoretical.  The possible
importance of these effects is stated in Ref.~\cite{kane95:um-th-95-16},
which finds that by including light superpartners in the analysis
of the LEP data, both effects can be simultaneously explained.
In particular, it is remarkable that the LEP $\alpha_s$ measurement
becomes consistent with other $\alpha_s$ measurements
(at $\alpha_s \approx 0.112$) when light charginos and stops
are included.  Further, Ref.~\cite{kane95:um-th-95-16} reports
a global analysis of the LEP/SLC data to ensure that no other
observable is adversely affected by the light superpartner
contribution (such as $\Delta\rho$ or $m_W$).  One finds
the supersymmetry fit is actually better than the SM fit,
giving better agreement to, for example, the $R_b$ and $A_{LR}$
measurements.  The better agreement requires both $m_{\charI}$
and $m_{{\tilde t}_1}$ to be less than about 100 GeV\@.

Surprisingly, the properties of the chargino and stop are sufficiently
well determined by the above analysis that one is forced into concluding
that the theoretical assumptions of the CMSSM cannot be fully correct.
The required mixture~\cite{wells94:219} of gaugino and Higgsino
in the chargino, and the left--right mixing for the stop, are not allowed
by the CMSSM, though they are fully allowed by the theory when particular
assumptions about parameters in the CMSSM are relaxed.  There has not been
time to construct constrained models based on the non-minimal theory,
but we can summarize the expected impact on the opportunity
to detect superpartners at the Fermilab Tevatron.

First, and most important, the light chargino and stop should
be fully produced at the Fermilab Tevatron.  The stop production
is predominantly via gluons and is unaffected since the cross section is
independent of $\tilde t_L$, $\tilde t_R$ mixing.
The chargino production rate will also not be greatly affected
over the parameter space.  Kinematically, since charginos and stops
are lighter than about 100 GeV, there is no suppression at the
Fermilab Tevatron.

The signatures and detection will be different, and can be more
difficult.  The analysis for stop production and detection
is largely the same.  However, charginos that resolve the $\alpha_s$
``crisis'' are more Higgsino--like than CMSSM charginos, which implies
the LSP mass is closer to the chargino mass, and so the resulting
leptons are softer (hence fewer leptons pass the $p_T$ cut).
One new helpful feature does enter, in that now all stops are lighter
than tops, so detecting stops in top decay is a major
opportunity.  The Br($t \ra {\tilde t}$ + LSP) has a factor
of $G_F m_t^3$ and the usual phase space factor, plus a
factor of $N_{i4}^2 / \sin^2 \beta$ that is the supersymmetric
modification from the Higgsino wave functions and the top--stop
coupling.  Numerically, we find
$0.15 \lsim$ Br($t \ra {\tilde t}$ + LSP) $\lsim 0.6$ for
$m_{\charI} \sim 60$ GeV, to
$0.05 \lsim$ Br($t \ra {\tilde t}$ + LSP) $\lsim 0.3$ for
$m_{\charI} \sim 90$ GeV.

Once the decay $t \ra {\tilde t}$ + LSP has occurred, the analysis
of $\tilde t$ decays is as before, with ${\tilde t}_1 \ra \charIplus b$
dominating if it is open, and ${\tilde t}_1 \ra \neutI c$ otherwise.
For chargino detection a new analysis is required to correctly
calculate the BR $\times$ EFF for the trilepton signal, and the needed
models have not yet been analyzed.  Qualitatively, as remarked above,
the leptons will be a little softer, so lowering the lepton triggers
will increase the efficiency, and gains by going to softer leptons may
be major.  The interference effect and the invisible $\snu\nubar$
mode may not cause such a large suppression of the leptonic branching
ratio of the neutralino as in the CMSSM, possibly improving the
situation for some of the parameter space.

In summary, if the LEP $R_b$ measurement and the
$\alpha_s$ crisis are telling us that light charginos and stops
exist, then they are already being produced at the Fermilab Tevatron.
The challenge is to detect them.  The analysis based
on the CMSSM, or an extended theory, can help thinking about
signature and detection efficiency, and help the interpretation if
no signal is found.  If light charginos and stops are there,
high luminosity and a careful search is needed to find them.

\section{Conclusions and Comments}
\indent

We have conducted a thorough study of many possible
supersymmetry signals at a luminosity upgraded Fermilab
Tevatron collider.  Our studies have complemented and often
improved on others using our full {\em event}--level Monte Carlo
based on {\sc Pythia/Jetset} coupled with a comprehensive approach
to constrained minimal supersymmetric parameter space.
We conclude, consistent with earlier
work, that the trilepton signal and gluino production are the
most useful signals to discover supersymmetry at the
Fermilab Tevatron collider.
Although no limits could be placed on the mass of the lightest chargino
if no signal were detected, the discovery potential at Fermilab
is quite large,
extending to chargino masses well beyond that which can be
probed by LEP~II\@.  With integrated luminosities of
200 ${\rm pb}^{-1}$, 2 ${\rm fb}^{-1}$ and 25 ${\rm fb}^{-1}$,
the Fermilab Tevatron can reach to chargino masses of
140 GeV, 210 GeV and 240 GeV respectively.
Gluino searches at Fermilab will be able to detect
solutions with $m_{\tilde g}$ up to $\sim$300 GeV, $\sim$350 GeV, and
$\sim$400 GeV to $10\sigma$ significance with integrated
luminosities of 200 ${\rm pb}^{-1}$, 2 ${\rm fb}^{-1}$, and
25 ${\rm fb}^{-1}$.  The stop production signal at Fermilab
has a stop mass reach up to $\sim$160 GeV at $5\sigma$ significance
in the $W$ + jets channel with an integrated luminosity of 2 ${\rm fb}^{-1}$.
At an integrated luminosity of 25 ${\rm fb}^{-1}$, Fermilab has a
stop mass reach up to $\sim$200 GeV at $5\sigma$ significance for
stop production in both the $W$ + jets and the dilepton channels.

While searches at LEP can likely find or exclude chargino and
stop masses up to nearly the beam energy $\frac{\sqrt{s}}{2}$,
searches at the Fermilab Tevatron have a reach $\sim$2--3 times
that of LEP~II\@.  But, the Fermilab Tevatron cannot set mass limits
if no signal is found because sets of parameters exist that give
smaller $\sigbreff$ than can be detected.  However, in the
absence of a discovery what matters is establishing tighter
experimental constraints on the parameters; in Section 3.3
we showed how a variety of
information will combine to reduce the allowed parameter space,
so that increasingly unique and testable predictions can be made.
In this pursuit, LEP~II and Fermilab are comparably powerful and
somewhat complementary.  Searches for supersymmetry utilizing
existing and upgraded collider facilities have a great
potential for discovery, and if no discovery occurs at
a given energy or luminosity then the results help sharpen both
our understanding of the theory and predictions for a variety
of other experiments.

\vspace{3cm}

\section*{Acknowledgments}

We would like to thank K.~De, C.~Kolda, S.~Martin
and R.~Watkins for helpful conversations.  We are indebted
to C.~Kolda for the use of his RGE program.  S.M\@. would like to
thank B.~Barish for support.  This work was supported
in part by the U.S.~Department of Energy.

\newpage



\begin{thebibliography}{99}
\bibitem{SUSYSM}
    We note that a Monte Carlo ``SUSYSM'' was developed
    for hadron supercolliders using an interface to the
    {\sc Pythia}/{\sc Jetset} system in Ref.~\cite{baer92:142}.
\end{thebibliography}
\end{document}